\documentclass{article}
\usepackage[T1]{fontenc} 
\usepackage[utf8]{inputenc} 
\usepackage{ismir,amsmath,cite,url}
\usepackage{graphicx}

\usepackage{color}

\usepackage[firstpage]{draftwatermark}
\definecolor{lightgray}{rgb}{0.9,0.9,0.9}
\definecolor{darkgray}{rgb}{0.4,0.4,0.4}
\SetWatermarkFontSize{12pt}
\SetWatermarkScale{1.1}
\SetWatermarkAngle{90}
\SetWatermarkHorCenter{202mm}
\SetWatermarkVerCenter{170mm}
\SetWatermarkColor{darkgray}
\SetWatermarkText{{Late-Breaking / Demo Session Extended Abstract, ISMIR 2022 Conference}}

\def\lbd{}	

\title{MIDI-Draw: Sketching to Control Melody Generation}

\multauthor
{Tashi Namgyal \hspace{1cm} Raul Santos-Rodriguez \hspace{1cm} Peter Flach} 
{ Department of Computer Science, University of Bristol, Bristol, UK\\
{\tt tashi.namgyal@bristol.ac.uk}
}

\def\authorname{T. Namgyal, R. Santos-Rodriguez and P. Flach}

\begin{document}

\maketitle
\begin{abstract}
    We describe a proof-of-principle implementation of a system for drawing melodies that abstracts away from a note-level input representation via melodic contours. The aim is to allow users to express their musical intentions without requiring prior knowledge of how notes fit together melodiously. Current approaches to controllable melody generation often require users to choose parameters that are static across a whole sequence, via buttons or sliders. In contrast, our method allows users to quickly specify how parameters should change over time by drawing a contour. 

\end{abstract}
\section{Introduction}\label{sec:introduction}

Recent deep learning approaches offer more variety, musicality and longer range structure than many previous approaches to music generation \cite{briot_deep_2020}. However, the high expressivity and black-box nature of these models makes it hard to predict what outputs a model will produce given certain inputs. Some understanding of the model is necessary for musicians to be able to iteratively tweak the model to produce outputs that better match their intentions, which is an important part of musical composition \cite{garcia_structured_2014}. This is referred to as \emph{controllability} or \emph{directability}. 

Many musical features have been targeted for control including low-level musical features such as key, tempo or mode \cite{meade_exploring_2019}, mid-level perceptual features such as tonal tension \cite{guo_variational_2020} or information-theoretic surprise \cite{chen_surprisenet_2021}, or high-level features such as artist style and lyrics \cite{dhariwal_jukebox_2020}. This is often achieved through conditional models, which change the probability of output sequences given some user-chosen feature values \cite{meade_exploring_2019}, or by manipulation of a latent space via attribute vector arithmetic \cite{roberts_hierarchical_2019} or attribute regularization and disentanglement \cite{pati_dmelodies_2020}. Control over these features enables users to feel more self-efficacy and ownership over the generated music \cite{louie_novice-ai_2020}.

Controls for generative models are most often presented as switches or 1D sliders \cite{louie_novice-ai_2020}. 2D sliders are also common, often displayed over an image such as a surface plot \cite{bryan-kinns_exploring_2021}. Other control methods include tracking body movements in space, known as sound tracing \cite{godoy_body_2009} and conditioning on text using contrastive language-image pre-training (CLIP) on videos with associated music \cite{wu_wav2clip_2022}. Several approaches exist that use drawing such as Oramics \cite{richards_oramics_2018}, Painting Music \cite{starkey_painting_2020}, Hyperscore \cite{farbood_hyperscore_2004} and JamSketch \cite{kitahara_jamsketch_2018}. Our proposal, MIDI-Draw, adopts this approach because drawing is a quick, easy and natural movement that does not need specialist equipment. A quick sketch is useful as it represents a semi-formed idea, which fixes some attributes while allowing freedom for others to change more easily. This allows for an intermediate representation that is intuitive to humans but that machine learning approaches are also sensitive to. 



\begin{figure}
    \centerline{\framebox{\noindent
    \includegraphics[width=0.7\columnwidth]{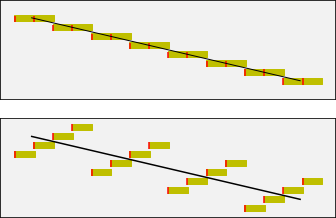}}}
    \setlength{\belowcaptionskip}{-8pt}
    \caption{Musical phrases can be generated with the same overall trend while allowing for variation in tightness of fit.}
    \label{fig:cloudiness}
\end{figure}

We demonstrate the potential of the approach on melodic contours, which capture how pitch goes up and down over time, while disregarding quantized note and scale information. Geometric approaches appear to be the most promising approach for symbolic melody indexing \cite{parsons_directory_2008} and similarity \cite{velardo_symbolic_2016}. Such methods include rectilinear functions \cite{aloupis_algorithms_2006}, cubic splines \cite{urbano_mirex_2013}, Fourier components \cite{kitahara_jamsketch_2018}, or cosine components \cite{cornelissen_cosine_2021}. Many of these approaches relate pitches directly to the curve height. We adopt cosine components here as they can extract a slow-moving average trend, which allows for more freedom and variety in generation. For example, notes may lie directly on the curve, or scattered around the curve while still fitting the trend overall (Fig. \ref{fig:cloudiness}).

\section{Methods}

\begin{figure}
    \centerline{\framebox{\noindent
    \includegraphics[width=0.8\columnwidth]{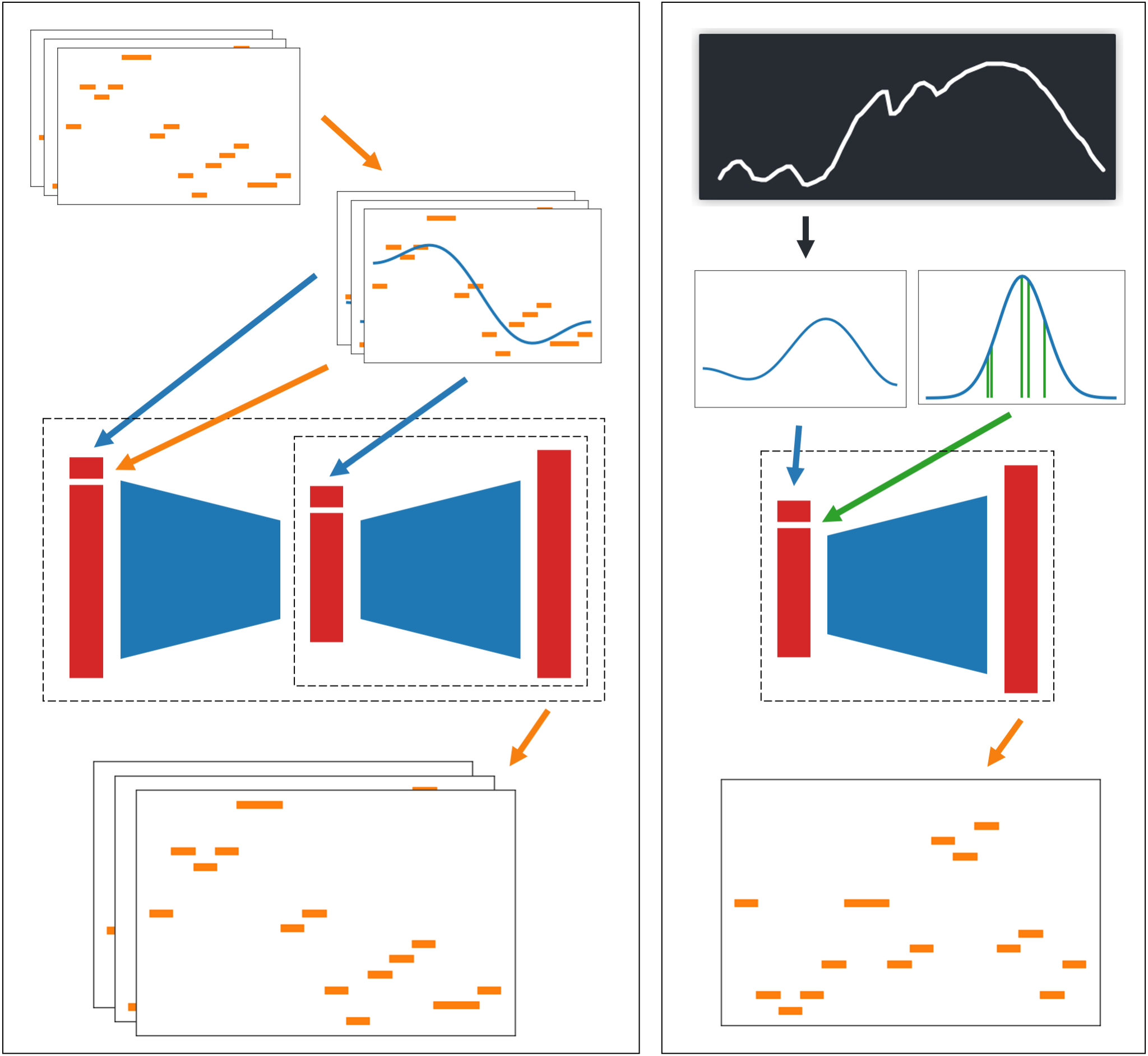}}}
    \setlength{\belowcaptionskip}{-8pt}
    \caption{Left: Training consists of extracting cosine components from melodies and recreating melodies conditioned on those components. Right: Inference consists of generating melodies by sampling from a prior distribution conditioned on cosine components extracted from a drawn curve.}
    \label{fig:systemdiagram}
\end{figure}

To train MIDI-Draw (Fig. \ref{fig:systemdiagram}, left), we use a synthetic dataset generated by specifying a pitch-transition matrix with uniform pitch priors. 5000 sequences of 16 notes are sampled from this matrix. These pitch sequences are converted into cosine component amplitudes using the orthonormal Discrete Cosine Transform (DCT-II). The average pitch of each melody (first component) and the higher frequency components are set to zero such that only the second to fourth cosine components are stored. This results in a low frequency (slow changing) pitch trend when passed back through the inverse (DCT-III) transform.

    
In order to generate music from a drawn shape rather than by continuing some musical context, we use a conditional variational autoencoder (CVAE). A VAE has two training objectives; to reconstruct the input and to maintain the latent space distribution close to a chosen prior distribution. The latter allows for sampling directly in the latent space during inference, without needing to encode some previous context. The CVAE is trained to recreate melodies conditioned on the cosine components by concatenating them to the start of the input and latent vectors. Bi-directional long-short term memory (Bi-LSTM) models are used for the encoder and decoder, which allow for long sequences and reuse of past context due to their hidden cell state. For the decoder, an autoregressive conductor model is used as in \cite{roberts_hierarchical_2019}. 
    

During inference (Fig. \ref{fig:systemdiagram}, right), a drawn curve is passed through the DCT-II transform as above, which is repeatedly concatenated to a batch of latent vectors generated by randomly sampling from the chosen prior, a multivariate Gaussian distribution $\epsilon \sim \mathcal{N}(0,I)$. This is then passed through the decoder to generate a batch of melody candidates. The DCT transform is again performed on the resulting melody candidates and mean-squared error is used to pick the melody candidate with components closest to the drawn curve components.
    
\section{Results and Discussion}
    
\begin{figure}
    \centerline{\framebox{\noindent
    \includegraphics[width=0.65\columnwidth]{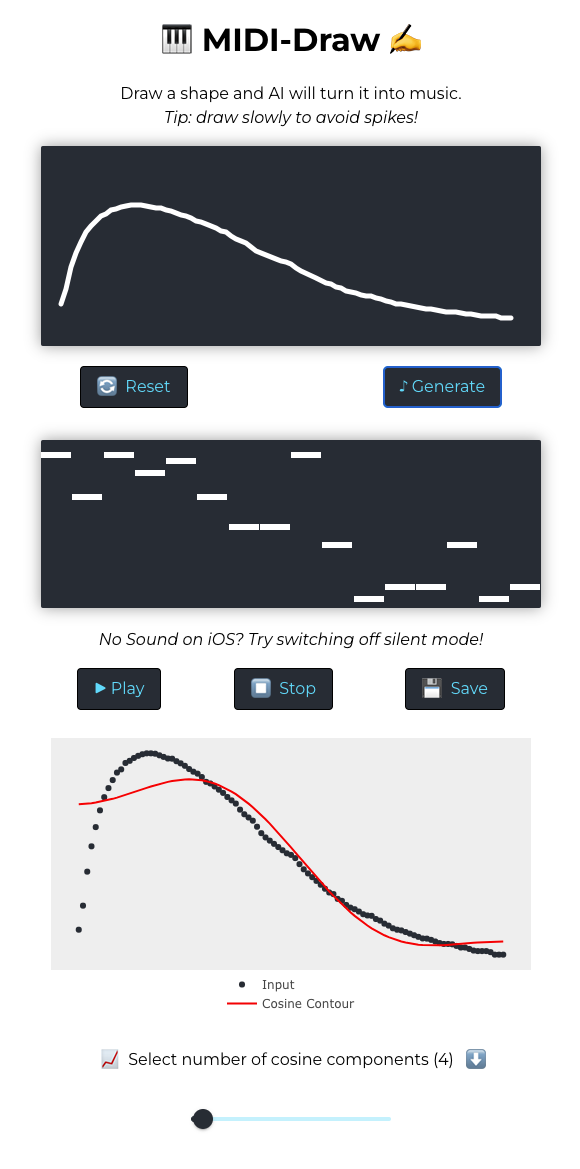}}}
    \setlength{\belowcaptionskip}{-8pt}
    \caption{Web app$^1$.}
    \label{fig:webapp}
\end{figure}
    
A web app\footnote{Web app available at \url{www.tashinamgyal.com/midi-draw}} (Fig. \ref{fig:webapp}) provides a canvas interface for users to draw contours and a piano roll interface to visualize, play back and download generated melodies. Preliminary user tests showed that non-musicians often initially drew curves that oscillated very rapidly, even though most melodies do not do this, suggesting additional constraints or guidance would be useful. A graph was added to show how changing the number of cosine components used would change how closely the generated melody fits their sketch, but further work is needed to train a model with an adjustable number of conditional components. Users also suggested that a constant rhythm made generated melodies sound robotic. Variation in rhythm could be added by conditioning on rhythm onsets\cite{chen_music_2020}, rhythm circles \cite{hein_groove_2019} or by tracing deviation from a regular beat. Lastly, we plan to work on evaluating the quality of musical sequences and how well sequences match user intentions. 


\section{Conclusion}

We described a proof-of-principle implementation of music generation from drawing. This frees novice musicians to comparatively, rather than absolutely, express musical intentions, while a generative model does the hard part of selecting notes that loosely yet melodiously follow the intended trend. We demonstrate the potential of the approach with a web app that allows users to draw pitch contours and generates matching melodies. This opens the way to user-centred music generation that tracks changes in musical intention over time more naturally than current alternatives.


\section{Acknowledgments}
Tashi Namgyal is supported by the UKRI Centre for Doctoral Training in Interactive Artificial Intelligence (EP/S022937/1).


%
%
%
%
%

\end{document}